\documentclass[aps,plb,reprint,amsmath,amssymb]{revtex4-1}

\usepackage{graphicx}
\usepackage{dcolumn}
\usepackage{bm}
\usepackage{amsmath, amssymb, mathtools}
\usepackage{tikz}
\usetikzlibrary{calc,decorations.markings}

\begin{document}

\title{AdS from Entanglement Entropy}

\author{Seungjoon Hyun}
\email{sjhyun@yonsei.ac.kr}
\author{Sang-A Park}
\email{sapark@kias.re.kr}
\affiliation{$^*$Department of Physics, College of Science, Yonsei University, Seoul 03722, Korea\\
$^\dagger$School of Physics, Korea Institute for Advanced Study, Seoul 02455, Korea}

\date{\today}

\begin{abstract}
We show that the anti-de Sitter(AdS) space naturally emerges from the conformal field theory (CFT). The behavior of the leading divergent term in  the entanglement entropy implies the underlying AdS geometry. The coefficient of the leading divergent term is related to the radius of the AdS space.  All these are confirmed fully for the two dimensional CFTs.  We also give comments for the higher dimensional CFTs.
\end{abstract}

\maketitle

\section{Introduction}

Ever since the AdS/CFT correspondence has been proposed in \cite{Maldacena:1997re}, which realizes the
holographic principle \cite{tHooft:1993dmi,Susskind:1994vu} in the context of superstring theory,  various attempts have been made to reconstruct the geometry starting from quantum field theory. 
One of the most promising approach turns out to be the one using the entanglement entropy which reveals  the underlying quantum nature of the physical system. In the context of the AdS/CFT correspondence, holographic interpretation of the entanglement entropy of the CFT  was proposed in \cite{Ryu:2006bv} and derived in \cite{Casini:2011kv,Lewkowycz:2013nqa}. This led to the idea that  entanglement entropy may be useful in the reconstruction of the corresponding geometry. Indeed, it has been argued  in \cite{VanRaamsdonk:2009ar}
 that quantum entanglement plays an essential role in the emergence of a dual geometry. 
It was indicated in \cite{Swingle:2009bg} that the tensor network methods  in the lattice system,  with particular emphasis on the role of entanglement renormalization, would generate  the discrete version of the spatial geometry of the AdS space. It follows that in \cite{Nozaki:2012zj} the metric in the multi-scale entanglement renormalization ansatz \cite{Vidal:2007hda} and its continuum version\cite{Haegeman:2011uy} was proposed and shown to be consistent with the AdS/CFT correspondence. For the additional developments in this direction, see  \cite{Beny:2011vh,Miyaji:2015yva,Pastawski:2015qua,Czech:2015qta,Miyaji:2015fia,Hayden:2016cfa,Caputa:2017urj}. Another interesting attempts to use holography for emergent gravity, see  \cite{Lee:2009ij}.

The aim of this paper is to show that the AdS spacetime naturally  emerges from the CFT by using the entanglement entropy in its simplest manner. In particular, it is shown that the behavior of the leading ultraviolet(UV) divergent term in the entanglement entropy
requires the emergence of the asymptotically AdS space and its coefficient represents the AdS radius.  In this sense, one may easily notice that the treatment of the two dimensional CFT is more robust and reliable compared to the one of the higher dimensional CFT.  Namely, the leading divergence of the entanglement entropy in two dimensional CFT is logarithmic and thus it's coefficient is universal, reflecting the characteristics of the corresponding CFT.
On the other hand, the leading divergent term in the  entanglement entropy in higher (d+1)-dimensional CFT behaves like $a^{-(d-1)}$ for the UV cut-off $a$ and thus it's coefficient is not universal.
Henceforth we focus mainly on the two dimensional CFT in various states, while  giving  short comments on the higher dimensional case.

We would like to view the entanglement entropy as an extensive quantity of the surface in the hidden space with one extra dimension which is homologous to the subsystem A and sharing the boundary, which is the boundary surface $\partial A$ dividing subsystem A and its complement of the CFT system.
Namely, we assume that the entanglement entropy is given by
\begin{equation}
S_A={{\rm Area} \over 4G_{N}}~,
\end{equation}
where the `Area' is the area of the extremal surface whose boundary is $\partial A$ and $G_{N}$ is a Newton constant in the hidden spacetime.  The coefficient has been chosen following the Bekenstein-Hawking entropy formula. Obviously, in order to have this kind of geometric description for the entanglement entropy, the area should be much larger than the Newton constant.  
The second assumption is that the UV cut-off or lattice spacing $a$ of the CFT comes from the cut-off of the hidden dimension. This means that the hidden direction represents, more or less, the energy scale of the CFT.  
Under these two assumptions, by using the leading order term in the entanglement entropy we show that the hidden spatial section of the AdS  geometry naturally emerges. The full AdS geometry would be recovered by the conformal symmetry.

\section{Infinite line and the Poincar\'e AdS$_3$}\label{section2} 
In this section we consider 
the entanglement entropy of the (1+1)-dimensional CFT on the infinite line  which is given by
\begin{equation}
S_A= {c\over 3} \log {\l \over a},
\end{equation}
with the central charge $c$ of the CFT.  The length scale $a$ denotes the lattice spacing or the UV cut off and the length $l$ does the length of the subsystem A.
In order for the notational simplicity, let us introduce a length scale defined as
\begin{equation}
R \equiv { 2 cG_N \over 3},
\label{R}
\end{equation}
in which
 the entanglement entropy can be expressed as 
 \begin{equation}
S_A={R \over 4G_N} \log \left({\l \over a}\right)^{2}.
\end{equation}
In order to have this geometric description of the CFT, the length scale $R$ should be much larger than the Newton constant $G_{N}$, i.e. the central charge $c$, which represents the degrees of freedom in the CFT, should be large.

Now we would like to consider this entanglement as the length of the line whose end points are the points separating the subsystem from its complement and whose bulk parts are located in a space which should be identified. 
We denote the coordinate along the hidden direction  as $z$, which is  orthogonal to the boundary spatial  coordinate $x$ where the CFT is on.  See Fig.\ref{fig:line}. 
The metric can be generally written as
\begin{equation}
ds^2=e^{-2\phi(z)}(dz^2+dx^2),
\label{metric}
\end{equation}
where the metric function $\phi(z)$ depends only on the $z$-coordinate due to the Poincar\'e invariance of the CFT.

\begin{figure}[b]
\begin{tikzpicture}[scale=.5]
\draw [->](-5,0) -- (5,0) node[right] {$x$};
\draw [->](0,0) -- (0,5) node[above] {$z$};
\draw[dashed] (-5,0.5) -- (5,0.5);
%
\draw[red,thick](4,0) arc(0:180:4);
\node at (-4,-0.5) {$-{l\over 2}$};
\node at (4,-0.5) {${l \over 2}$};
\node at (-0.3,0.8) {$a$};
\node at (-0.3,4.2) {$z_*$};
\end{tikzpicture}
\caption{\label{fig:line}}
\end{figure}

One dimensional geodesic equation in this geometry is the Euler-Lagrange equation of the action of the curve,
\begin{eqnarray}
I= \int^{\frac{l}{2}-\epsilon}_{-\frac{l}{2}+\epsilon}dx \;e^{-\phi(z)}\sqrt{\dot{z}^2+1}, 
\end{eqnarray}
where the geodesic line is parametrized by the boundary coordinate $x$. The cut-off of the $x$ coordinate is denoted as $\epsilon$ and  the derivative is the one with respect to the parameter of the curve, $\dot{z}\equiv{dz\over dx}$.
Our basic strategy is to determine the metric $e^{-2\phi}$, which plays the role as the potential of the action, from the integrated result.
The geodesic equation can be solved by noting that 
the action  is  invariant under translation along $x$-direction, which plays the role of time in this classical mechanical problem. 
Therefore the corresponding Hamiltonian, which is conserved along the curve, is given by
\begin{equation}
H= -{e^{-\phi(z)}\over \sqrt{\dot{z}^2+1}}\equiv - e_0,
\label{solution1}
\end{equation}
and $e_0$ is a constant along the geodesics. 
Because of the invariance under $x \rightarrow -x$,  $\dot{z}$ should vanish as $x$ approaches the origin $0$, i.e., turning point. 
If  the $z$ coordinate of this turning point  is denoted as $z(0)\equiv z_{*}$ and $\phi(z)$ at that point as  $\phi_{*}\equiv\phi(z_{*})$,
the value of Hamiltonian is given by $-e_{0}=-e^{-\phi_{*}}$.

After plugging this back to the action, the on-shell action should be identified as the geodesic length as
\begin{eqnarray}
I=   e^{\phi_{*}}\int^{\frac{l}{2}-\epsilon}_{-\frac{l}{2}+\epsilon}dx \;e^{-2\phi(z(x))}= R \log \left({l \over a}\right)^{2}.\label{action1-2} 
\end{eqnarray}
The coordinates of the end of the geodesic line is given by $(z, x)=(0, \pm {l\over 2})$. Toward these end points the length becomes logarithmically divergent and we need to introduce the cut-off. 

As mentioned earlier, we assume the cut-off of the hidden direction $z$ as $a$. 
What would be the corresponding cut-off $\epsilon$ of the coordinate $x$? Note that at the end points, $\dot{z}$ becomes divergent. Therefore near the end points,  we have
$
\dot{z} \simeq  e^{\phi_*-\phi}
$
from the equation (\ref{solution1}),
which  gives the relation
\begin{eqnarray}
\epsilon =\int^{-\frac{l}{2}+\epsilon}_{-\frac{l}{2}}dx=e^{-\phi_{*}}\int^{a}_{0}dz\, e^{\phi}, \label{epsilon}
\end{eqnarray}
for the infinitesimal $a$ and $\epsilon$. 
 On the other hand, by differentiating the equation in (\ref{action1-2}) with respect to $a$, we have
\begin{eqnarray}
-2 e^{\phi_{*}} e^{-2\phi(a)}\frac{d\epsilon}{da} =-{2R\over a}.
\label{epsilon1}
\end{eqnarray}
By comparing (\ref{epsilon}) and (\ref{epsilon1}), we find the metric function at the cut-off as $e^{\phi(a)}=\frac{a}{R} $ and the relation between those two cut-off as $\epsilon={a^{2}\over 2 e^{\phi_{*}}R}$.

Now we are in position to deal with the integration (\ref{action1-2}). 
First of all, let us note that the metric  $e^{-2\phi(z)}$ as a function of the parameter $x$ along the geodesics diverges as $x\rightarrow \pm \frac{l}{2}$  and, from the logarithmic dependence of the integrated length on the cut-off $\epsilon$,  it is clear that $x= \pm \frac{l}{2}$ should be simple poles.
Furthermore the geodesic line must be symmetric under $x\rightarrow -x$, reflecting the rotational invariance of the underlying CFT.  
Therefore the metric can be generally expressed as  
\begin{equation}
e^{-2\phi(z)}= A_{0}\left({1\over \frac{l}{2}+x}+{1\over \frac{l}{2}-x}\right)+F_0(x),
\label{phi}
\end{equation}
where $A_{0}$ is  a constant to be determined and $F_0(x)$ is a regular function in the region $-\frac{l}{2}\leq x \leq \frac{l}{2}$. 

The regular function $F_0(x)$ can be generally expressed as a Taylor series in terms of $\frac{2x}{l}$, whose integral is given by the power series of  $l$.
On the other hand, by plugging (\ref{phi}) back to (\ref{action1-2}), the value of the action is given by 
\begin{eqnarray}
I
&=& R\log \left({l \over \epsilon}\right)+ R\log \left({l \over 2e^{\phi_{*}}R}\right).
\end{eqnarray}
The first term should come from the simple poles and thus require $ A_{0} e^{\phi_{*}}=\frac{R}{2}$, 
while the second term should come from the integration of the regular function $F_0(x)$.

However, since it is not possible to have $\log l$ nor a constant by the definite integral of the polynomials of $x$, the second term should vanish and therefore  the metric function at the turning point should be given by $e^{\phi_{*}}=\frac{l}{2R}$. This determines  the coefficient of the simple pole as $A_{0}={R^{2}\over l}$ and  
the metric as
\begin{equation}
e^{-2\phi(z)}=\frac{R^2}{l}\left({1\over \frac{l}{2}+x}+{1\over \frac{l}{2}-x}\right).
\end{equation}


Now it is straightforward to find out the metric of the geometry. Since the derivative $\dot{z}$ satisfies 
\begin{equation}
\dot{z}^2+1={l\over 4}\left({1\over \frac{l}{2}+x}+{1\over \frac{l}{2}-x}\right),
\end{equation}
 from the eq. (\ref{solution1}), the equation governing the geodesic curve is easily found to be
\begin{equation}
	z=\sqrt{\frac{l^2}{4}-x^2 },
\end{equation}
and the metric function is determined as
\begin{equation}
	e^{\phi(z)}=\frac{z}{R}.
\end{equation}

Therefore the  metric turns out to be the one of the spatial part of the  AdS geometry in Poincar\'e coordinates as
\begin{equation}
ds^2=\frac{R^2}{z^2} (dz^2+dx^2).
\end{equation}
The whole AdS geometry including time naturally emerges by considering the conformal symmetry which is reflected as the AdS isometry.
The length scale $R$ is identified as the radius of the AdS space and the
central charge $c$ of the CFT is related to the AdS radius  as 
\begin{equation}
c={3R \over 2G_N},
\end{equation}
which is the famous Brown-Henneaux relation\cite{Brown:1986nw}.

\section{Circle and the Global AdS$_3$}
In this section we would like to consider the entanglement entropy of the subsystem A in a circle.  
We show that the corresponding emergent geometry is the AdS$_{3}$ in global coordinates.
The entanglement entropy of the (1+1)-dimensional CFT on the circle is well-known to be
\begin{equation}
S_A= {c\over 3} \log \left({L \over \pi a}\sin{\pi l\over L}\right)\,,
\end{equation}
where $ L$ denotes the length of the circle and $l$ does the length of the subsystem A.
We use the same length scale $R$ as in the previous section and express  the entanglement entropy  as
 \begin{equation}
S_A={R \over 4G_N} \log \left({L \over \pi a}\sin{\pi l\over L}\right)^{2}.
\end{equation}

It is straightforward to  find out the geometry behind this entanglement entropy by following the same logic as in the previous section.
Firstly, we replace the boundary coordinate $x$ by the compact coordinate ${L\over 2\pi}\theta$ in the metric, with $\theta\sim \theta+ 2\pi$.
 The action of the curve is invariant under the translation in the parameter $\theta$ and the corresponding conserved Hamiltonian is given by   
\begin{equation}
H = -L_0^2{e^{-\phi(z)}\over \sqrt{z'^2+L_0^2}}= - L_0e^{-\phi_*},
\label{solution2}
\end{equation}
where the $\theta$-derivative is denoted as $z'\equiv {dz\over d\theta}$ and the same notation is used for the turning point,  $z(0)=z_{*}$ and $\phi_{*}=\phi(z_{*})$. 
The resultant action becomes
 \begin{eqnarray}
I=   L_{0}e^{\phi_{*}}\int^{\theta_{0}-\epsilon}_{-\theta_{0}+\epsilon}d\theta\, e^{-2\phi(z(\theta))}.\label{action2-2} 
\end{eqnarray}
where the coordinate at the end point is denoted as $\theta_{0}={\pi l\over L}$.

By using the same method to determine $\epsilon$ as in the case of infinite line, we find  that $e^{\phi(a)}={a \over R} $ and  $\epsilon={\pi \over LRe^{\phi_{*}}}a^{2}.$
The metric $e^{-2\phi}$ along the geodesic line, expressed in terms of the parameter $\theta$, is  periodic under $\theta\rightarrow \theta+2\pi$ and symmetric under $\theta\rightarrow -\theta$. Therefore it  consists of singular terms with  simple poles at $\theta=\pm \theta_{0}$ and  the regular terms as
\begin{equation}
e^{-2\phi(z)}
= A_{1}\left({1\over \sin\theta_{0}+ \sin\theta}+{1\over \sin\theta_{0}- \sin\theta}\right)+F_1(\theta),
\label{phi2}
\end{equation}
where $A_{1}$ is a constant and $F_{1}$ is the regular part which would be  expanded as the Fourier series.  

Once again, by plugging (\ref{phi2}) back to the on-shell action (\ref{action2-2}),
we find  
$${L e^{\phi_{*}} A_{1}\over \pi\cos\theta_{0}}=R,$$
from the integration of the singular part.
On the other hand, if we take 
\[
e^{\phi_{*}}=\frac{L}{2\pi R}\tan{\theta_{0}},\label{value2}
\]
then the integration of the regular part vanishes, which means $F_1=0$. Therefore  the metric along the geodesic is given by
\begin{equation}
e^{-2\phi(z)}=\frac{2\pi^{2} R^{2}\cos\theta_{0}}{L^{2}\tan{\theta_{0}}}\left({1\over \sin\theta_{0}+ \sin\theta}+{1\over \sin\theta_{0}- \sin\theta}\right).
\end{equation}

Now we can determine the metric as a function of $z$. From the Hamiltonian (\ref{solution2}), 
 the equation of curve is given by
\begin{equation}
	z =\frac{L}{2\pi}\cosh^{-1}\left(\frac{\cos\theta}{\cos\theta_0}\right),
\end{equation}
which determines the metric function as
\begin{equation}
	e^{2\phi}=\frac{L^2}{4\pi^2 R^2}\sinh^2\left(\frac{2\pi z}{L} \right).
\end{equation} 
This confirms that  the spatial part of the global AdS$_{3}$ emerges from the entanglement entropy.  
Note that by introducing the  radial coordinate  $\rho$, which is related to the coordinate $z$ as  
$$\sinh \rho\sinh{2\pi z \over L}=1,$$  the metric  can be expressed as the standard form: 
\begin{equation}
	ds^2=R^2\left( d\rho^2 + \sinh^2\rho\, d\theta^2 \right).
\label{global metric}
\end{equation}

\section{Excited States and the AdS$_{3}$ with Deficit Angle}
In the previous sections, we have shown that the AdS$_{3}$ in the Poincar\'e and global coordinates emerges from the entanglement entropy of the CFT in the ground state on ${\bf R}$ and ${\bf S}^{1}$, respectively. Another interesting case is the entanglement entropy of the subsystem A in  the CFT in the excited  state.
It is known that the CFT in the excited state due to a twist operator with conformal dimension $\Delta={c \over 12}(1-{1\over n^{2}})$ is related to the AdS$_{3}$ with the deficit angle $\gamma\equiv n^{-1}$. 
The entanglement entropy in the background of this operator is known to increase by $\Delta S_{A}={2\Delta \over 3}\left({\pi l\over L}\right)^{2}+\cdots$ in the limit $l\ll L$, compared to the entanglement entropy in the vacuum state. In the holographic interpretation of the entanglement entropy, this increase of  the entanglement entropy in the excited state is described as coming from the limit $l\ll L$ of the entanglement entropy formula
\begin{equation}
S_A={2R \over 4G_N} \log \left({L \over \gamma \pi a}\sin{\gamma \pi l\over L}\right)~.
\end{equation}

If we accept the above formula as the entanglement entropy of the CFT in the excited state due to the above twist operator, then it is straightforward to obtain the AdS$_{3}$ geometry with deficit angle $\gamma$ by replacing $a\rightarrow \gamma a$ and $l\rightarrow \gamma l$ in the previous section. We obtain the same form of the metric as (\ref{global metric}) while the period of $\theta$ is $2\pi\gamma$. If we use the standard angle coordinate $\varphi={\theta\over\gamma}$ and radial coordinate $r=R\gamma \sinh\rho$, the metric becomes
\begin{equation}
ds^{2}=\frac{1}{{r^{2}\over R^{2}}+\gamma^{2}}dr^{2}+r^{2}d\varphi^{2},
\end{equation}
which is the spatial section of the AdS geometry with the deficit angle.

\section{Finite Temperature and the BTZ black holes}
In this section, we show that the BTZ black holes emerges from the entanglement entropy of the CFT in the finite temperature. For simplicity, we consider the entanglement entropy of a subsystem in the infinite line.
Obviously, it would give rise to planar BTZ black holes. The entanglement entropy for the thermal system with temperature $T=\beta^{-1}$ is known to be given by
\begin{equation}
S_A= {c\over 3} \log \left({\beta \over \pi a}\sinh{\pi l\over \beta}\right).
\end{equation}

We use the same form of the length scale and the metric as in (\ref{R}), (\ref{metric}). 
By considering the geodesic equation and the conserved Hamiltonian as in sec. II, we obtain the relation $\epsilon={ a^{2}\over  2Re^{\phi_{*}}}$ which leads to the on-shell action as
\begin{eqnarray}
I&=& e^{\phi_{*}}\int^{\frac{l}{2}-\epsilon}_{-\frac{l}{2}+\epsilon}dx \;e^{-2\phi(z)}\nonumber \\
&=& R\log  \left(\frac{\beta}{2\pi\epsilon} \sinh {2\pi l\over \beta}\right) +R\log \left(\frac{\beta\tanh{\pi l\over \beta}}{2\pi Re^{\phi_{*}}}\right).
\label{action3}
\end{eqnarray}

This time it would be better to expand the metric in terms of the hyperbolic function.
One may note that the metric has the singularities at $x=\pm\frac{l}{2}$ and symmetric under $x\rightarrow-x$. Then by using the same procedure as  in previous sections, one can determine the expression of the metric in terms of parameter $x$ as
\begin{align}
e^{-2\phi(z)}&=\frac{2\pi^{2} R^{2}\cosh{\pi l \over \beta}}{\beta^{2}\tanh{\pi l \over \beta}}\nonumber\\
&\quad \cdot\left({1\over \sinh{\pi l \over \beta} + \sinh{2\pi x \over \beta}}+{1\over \sinh{\pi l \over \beta} - \sinh{2\pi x \over \beta}}\right).
\label{phi3}
\end{align}
This gives us the curve equation as
\begin{equation}
	z =\frac{\beta}{2\pi}\cos^{-1}\left(\frac{\cosh{2\pi x\over\beta}}{\cosh{\pi l\over\beta}}\right),
\end{equation}
which determines the metric function in terms of $z$ as
\begin{equation}
	e^{2\phi}=\frac{\beta^2}{4\pi^2 R^2}\sin^2\left(\frac{2\pi z}{\beta} \right).
\end{equation}

By introducing $r_{+}\equiv {2\pi R^2\over \beta}$ and $r\equiv r_{+}\csc(\frac{2\pi z}{\beta})$, we obtain the spatial part of the BTZ geometry:
\begin{equation}
ds^{2}=\frac{R^{2}}{r^{2}-r_{+}^{2}}dr^{2}+{r^{2}\over R^2}dx^{2}.
\end{equation}

The computation of the entanglement entropy for the circle in finite temperature quantum field theory is more involved, but connecting it to the BTZ black holes is straightforward following the logic given in this section, for the case that the size of the entangling region $l$ is small enough  compared to the size of the total system.

\section{Higher Dimensional Cases}
In this section, we consider the entanglement entropy of the ($d+1$)-dimensional CFT, with $d\geq 2$. 
There exist some subtleties to apply our strategy to the higher dimensional cases. The  leading divergent term of the entanglement entropy in ($d$+1)-dimensional CFT is proportional to  $a^{-(d-1)}$ which means that  it's coefficient depends on the regularization and  is not universal. Nevertheless, in parallel to the (1+1)-dimensional CFT,  it can be shown that the existence of such a divergent term gives rise to the asymptotically AdS spacetime, whose radius is related to the coefficient of the leading divergence term.
For our purpose, it may enough to consider the simplest case, i.e., the entanglement entropy for the strip with the width $l$. The subsystem has ($d-1$)-dimensional translation invariance, whose coordinates are denoted as $x_{i}$, $i=2,\cdots d$, with IR regulator $L$.

Generically, the entanglement entropy in higher dimensions has the form \cite{Bombelli:1986rw,Srednicki:1993im}
\begin{equation}
S_{A}=c_{d-1} {{\rm Area}(\partial A) \over {a}^{d-1}}
+{\cal O}\left(a^{-(d-3)}\right),
\end{equation}
where $c_{d-1}$ is a constant which depends on the system and the structure of subleading terms depends on the intrinsic and extrinsic geometry of the entangling surface $\partial A$. In the case of strip, all the subleading divergent terms are absent due to the flatness of the entangling surface, while divergence-free term proportional to $({L\over l})^{d-1}$ is expected to exist. Henceforth the entanglement entropy of the strip may be written as
\begin{equation}
S_{A}=c_{d-1} \left({L \over a}\right)^{d-1}+c_{0} \left({L \over l}\right)^{d-1}.
\end{equation}
Now let us introduce a length scale $R$ which is related to the coefficient $c_{d-1}$ as  
$
c_{d-1}={2 R^{d} \over 4G_{N}(d-1)},
$
and try to find the  area of hidden surface bounded by the entangling surface $\partial A$ as
\begin{equation}
	A_{S}=\frac{2R^d}{d-1}\left(\left(\frac{L}{a}\right)^{d-1} +{c_{0}\over c_{d-1}} \left(\frac{L}{l} \right)^{d-1}\right).
\end{equation}

We are looking for the geometry with 
\begin{equation}
	ds^2=e^{-\phi(z)}\left(dz^2+dx^2+\sum_{i=2}^{d}dx_i^2 \right),
\end{equation}
where the metric function $\phi$ depends only on the $z$-coordinate. 
The area is given by an extremum of the action
\begin{equation}
	A_S=L^{d-1}\int_{-\frac{l}{2}+\epsilon}^{\frac{l}{2}-\epsilon} dx e^{-d\phi(z)}\sqrt{\dot{z}^2+1}, 
\end{equation}
which is invariant under the $x$-translation and the corresponding invariant quantity is given by 
\begin{equation}
{e^{-d\phi(z)}\over \sqrt{\dot{z}^2+1}}=e^{-d\phi_{*}} .
\label{solution4}
\end{equation}

By using the same method as in the case of infinite line in section \ref{section2},  the value of the metric at $z=a$ is given by $e^{-2\phi(a)}=({R \over a})^{2}$ and the relation between the cut-off in the $x$-coordinate $\epsilon$ and the cut-off in the $z$-coordinate is found to be
$$
\epsilon=\frac{a^{d+1}}{(d+1)e^{d\phi_*} R^d},
$$
in terms of which the area can be expressed as
\begin{eqnarray}
A_S=\frac{2R^{2d \over d+1}L^{d-1}}{(d-1)(d+1)^{d-1\over d+1}e^{{d(d-1)\over d+1}\phi_{*}}}\epsilon^{-{d-1\over d+1}} +\cdots.
\label{area4}
\end{eqnarray}
As mentioned earlier,
it is not easy to determine the metric function whose  integration gives the area, partly because of the ambiguity in the coefficient of the leading divergent term. 

Still we can find the behavior of the metric near the boundary cut-off as follows. 
The leading divergent term in the expression of area indicates that, as the geodesic surface approaches toward the boundary, i.e., $x\rightarrow \pm {l\over 2}$, the metric function along the geodesic surface diverges as $ e^{-2d\phi(z(x))}\sim \left( {l\over 2}\mp x\right)^{-{2d \over d+1}}$. This means that, near  $x= \pm {l\over 2}$, the metric function along the geodesic surface would be expressed as  
\begin{equation}
e^{-2d\phi(z(x))}\simeq A_{3}\left(\frac{l}{2}\mp x\right)^{-{2d\over d+1}}+F_{3}\left(\frac{l}{2}\mp x\right),
 \end{equation}
respectively, where $A_{3}$ a constant and $F_{3}(\frac{l}{2}\mp x)$ is some regular function of $\frac{l}{2}\mp x$. Then the area becomes
\begin{eqnarray}
A&=&L^{d-1}\int_{-\frac{l}{2}+\epsilon}^{\frac{l}{2}-\epsilon} e^{d\phi_{*}}e^{-2d\phi}dx\\
&=&\frac{2(d+1)}{(d-1)}L^{d-1}A_{3}e^{d\phi_{*}}\epsilon^{-\frac{d-1}{d+1}}+ {\rm regular\, terms},
\nonumber
\end{eqnarray}
whose leading divergent term should be identified with the one in eq. (\ref{area4}) and thus the coefficient $A_{3}$ is related to $e^{\phi_{*}}$ as 
\begin{eqnarray}
A_{3}=\left(\frac{R}{(d+1)e^{{d\phi_{*}}}}\right)^{2d \over d+1}.
\end{eqnarray}

On the other hand, Since $\dot{z}^{2}\gg 1$ near $x=\pm{l\over 2}$, the curve equation (\ref{solution4})  would be approximated as  
\begin{equation}
\dot{z}^{2}\simeq  e^{2d(\phi_*-\phi)}\simeq \left(\frac{Re^{\phi_{*}}}{d+1}\right)^{2d \over d+1}  \left(\frac{l}{2}\mp x\right)^{-{2d\over d+1}}, 
\end{equation}
near the boundary.
 This can be integrated to give the curve equation near the boundary as 
\begin{eqnarray}
z \simeq   \left((d+1)R^{d}e^{d\phi_{*}}\right)^{1 \over d+1} \left(\frac{l}{2}\mp x\right)^{1 \over d+1},
\end{eqnarray}
and thus the metric  is found to be 
\begin{equation}
e^{-2\phi}\simeq \left(\frac{R}{(d+1)e^{{d\phi_{*}}}} \frac{1}{\delta x} \right)^{2 \over d+1}\simeq \frac{R^{2}}{z^{2}},
\end{equation}
which clearly shows that it is asymptotically AdS space.

\section{Discussions}
In this paper we have shown that the AdS space naturally emerges from the entanglement entropy. 
We have made two assumptions: firstly,   
the entanglement entropy can be considered as an extensive quantity and thus related to the extremal volume which is located in the space with one extra dimension and is bounded by the surface $\partial A$ and secondly, 
the UV cut-off or lattice spacing  of the CFT comes from the IR cut-off of the hidden dimension. 
Then the  leading divergent term in the entanglement naturally implies the hidden AdS space and its coefficient is directly related to the AdS radius. In two dimensional CFT, this fully recovers all the background AdS geometry, including AdS with conical singularity and BTZ black holes. In higher dimensional CFT, it is more dubious because the coefficient of the leading divergent term, which is directly related to the AdS radius, is not universal. Nevertheless, the behavior of the leading divergent term guarantees the emergence of the AdS geometry.

In this work, only the spatial section of the AdS spacetime is recovered.  Even though the conformal symmetry would guarantee the emergence of the full AdS spacetime, still it would be nice if one can show that the full geometry can be recovered by extending our arguments.  It would be also very interesting to extend our work to the non-relativistic conformal field theory as well as quantum field theory away from critical point.
In any case it is very nice to see the emergence of the AdS geometry from the entanglement entropy in the simplest context.
\begin{acknowledgments}

We would like to thank Jae-Suk Park and Kyung Kiu Kim  for useful discussion. SH would like to thank KEK and KIAS for hospitality, where part of this work has been done. This work was supported by the National Research Foundation of Korea(NRF) grant with the grant number NRF-2016R1D1A1A09917598 and by the Yonsei University Future-leading Research Initiative of 2017(2017-22-0098).
\end{acknowledgments}
\appendix


\end{document}